\documentclass[numreferences]{kluwer}
\usepackage{klucite}
\usepackage[dvips]{graphicx}
\begin{document}
\begin{article}
\begin{opening}
\title{Color, spectral and morphological transformations of galaxies in clusters}            

\author{B.M. \surname{Poggianti}\email{poggianti@pd.astro.it}}
\institute{Padova Astronomical Observatory}                               


\begin{abstract} 
In this review I focus on the galactic properties in clusters at
$z>0.1-02$.  The most salient results regarding the evolution in
galaxy colors, spectral features and morphologies are discussed.
\end{abstract}




\end{opening}

\vspace{-0.8cm}
\section{Blue galaxies: the Butcher-Oemler effect}
The Butcher-Oemler effect is the excess of galaxies bluer than the
color-magnitude sequence (where most ellipticals/passive galaxies lie)
in clusters at $z>0.1-0.2$ as compared to the
richest nearby clusters \cite{bo78}. Butcher and Oemler were well
aware that the fraction of blue galaxies depends on a number of things, 
including the cluster type, the clustercentric radius considered and the galaxy
magnitude limit, as discussed below.\newline  
\noindent {\sl Optical versus X-ray cluster selection} -- The dependence of the
blue fraction on the cluster properties is potentially critical,
given that selecting different types of clusters at different
redshifts could in principle mimic an evolution.  For this reason, it is
interesting to ask whether optically and X-ray selected samples of
clusters reach similar conclusions regarding the Butcher-Oemler effect
\cite{smail98,andreon99}. Based on the
CNOC (Canadian Network for Observational Cosmology) 
X-ray-selected cluster sample, an evolution of the blue
fraction with redshift, very similar to the original Butcher-Oemler
result, has been confirmed \cite{kb01,elli01}.  Interestingly, the
blue fraction shows no simple trend with cluster X-ray luminosity
\cite{smail98,andreon99,fairley02}, a point I will discuss later.
\newline
{\sl Richness} -- How the blue fraction ($f_B$) depends on cluster richness 
has been the subject of an extensive study by \cite{margoniner01}, who
found $f_B$ to be higher in poor clusters and lower in rich clusters,
and to increase with redshift for all types of clusters.  It would be
important to verify how $f_B$ depends on richness using a
clustercentric radial limit that varies with the cluster scalelength,
instead of the fixed metric radius (0.7 Mpc) adopted by this study, 
because this choice could induce a spurious
trend with richness by sampling different areas in rich and
poor clusters \cite{fairley02}.
\newline 
{\sl Substructure} Possibly the most relevant question of all is how
the blue fraction, and the star formation activity in general, depend
on cluster substructure, i.e. on the 
merging/accretion history of the cluster.  The exact relation between
subcluster merging and blue fraction still needs to be quantified,
though there is a tendency for clusters with the highest blue
fractions to show signs of recent merging events
\cite{metevier00,wang97,smail98,pimbblet02}. Based on spectroscopy,
there is of course ample evidence for larger fractions of starforming
galaxies in substructured than in relaxed clusters. This trend
can originate from a higher average SF activity in the smaller
(less massive) systems that make up substructured clusters
\cite{abraham96,biviano97}, but can also be amplified by starbursts
induced by the merger itself \cite{moss00,cald93,metevier00,cald97}
as predicted
by several theoretical simulations.

Given the dependence of blue fraction on the cluster characteristics,
understanding how $f_B$ evolves at $z>0.5$ is more problematic
with the current cluster samples: a very massive cluster at z=0.8
(MS1054-03) has a blue fraction similar to typical values in clusters
at z=0.4-0.5 \cite{vd00}, while in other $z\sim 1$ clusters higher
fractions approaching 70-80\% have been reported
\cite{rakos95,tanaka00}.

Two aspects are worth highlighting here.  First, we should appreciate
that fully understanding (at any given redshift) how the galaxy
properties depend upon the cluster properties is very useful in order
to uncover how environment affects galaxy evolution.  Thus, the
dependence of the galactic properties on the cluster characteristics
is not a "burden" that is in our way when trying to disclose
evolutionary effects. On the contrary, it is in a sense the very thing
we are looking for.

Second, the lack of simple correlations between the galaxy properties
and the cluster most general (easier to observe) characteristics such
as the total X-ray luminosity or the optical richness shows that a
single global cluster property of this kind is an insufficient
information to characterize the cluster status in a useful way for
galaxy evolution studies.  More detailed pieces of information are
needed, including among others quantitative measurements of the
cluster mass, of the ICM local density, the knowledge of the dynamical
status of the cluster and its history of accretion of other
clusters/groups.  In this sense, the wavelength criterion for cluster
selection (optical versus X-ray) is not useful on its own, because
there is no cluster selection method that is immune from the problem
of carefully understanding a posteriori, on a cluster-by-cluster
basis, what is the cluster status and history.

Great care is required to compare $f_B$ values of different clusters
in a meaningful way also because the blue fraction depends on the
magnitude limit adopted, on the passband (\cite{fairley02}, see also
De Propris in these proceedings) and on the clustercentric
radius. \cite{elli01} have shown that when looking only at the central
core region (typically $\sim 0.5$ Mpc), no trend with redshift is observed,
while the Butcher-Oemler effect is conspicuous outside of the core.
The radial distribution of the blue galaxies and its evolution with z
must be related to the infall of field galaxies onto the cluster.  The
declining infall rate onto clusters at lower z is likely to play an
important role in the Butcher-Oemler effect, together with the
evolution of the average star formation rate in the field galaxies and
the decline of star formation in cluster galaxies \cite{elli01,kb01}. 
The Butcher-Oemler effect and the fact that the galaxy populations in 
clusters evolve
significantly during the last Gyrs are confirmed by the spectroscopic
and morphological studies described below.  This evolution involves
luminous giant galaxies, though probably not {\tt the most} massive
early-type galaxies that populate the top end of the magnitude
sequence, as discussed in the next section.
\vspace{-0.8cm}
\section{Red galaxies: the color-magnitude sequence}
There is a general consensus regarding the ages of the stellar populations 
in early-type cluster galaxies, i.e. in ellipticals and S0s as
morphologically-classified from HST images. The slope, zero-point and
scatter of their color-magnitude (CM) relation at various redshifts
indicate passive evolution of stellar populations that formed at
$z>$2-3, and the CM slope is found to be mostly driven by a
mass-metallicity relation
(\cite{ellis97,stanford98,kodama98,gladders98, vd00}, see also
\cite{bower92}). Moreover, well defined CM red sequences in higher redshift
clusters ($z \ge 1$) are now being observed
(e.g. \cite{stanford97,rosati99,lubin00,kajisawa00}), and there
are hints for a flatter CM sequence at high-z \cite{rosati01,vd01a}
that still needs to be comprehended.  The old stellar ages in
early-type galaxies are confirmed also by their spectroscopic features
(see below, e.g. \cite{postman98,p99}) and by fundamental-plane,
mass-to-light ratio and Mg-$\sigma$ studies
(e.g. \cite{vd96,kelson97,kelson00,kelson01,
bender96,ziegler97,ziegler01}), though this latter type of works have been
necessarily limited to relatively small numbers of the brightest
galaxies.

This homogeneity of histories of early-type galaxies in clusters
needs to be contrasted with a number of other results and considerations:

a) The population of early-type galaxies observed in distant
clusters does not necessarily comprises {\sl all} the early-type
galaxies existing at z=0. Morphological transformations might
occur in clusters (see below) and change later-type starforming galaxies
into some of the early-type galaxies present in clusters today.
Due to this ``progenitor bias'', in distant clusters 
we would be observing only those galaxies that were already assembled as 
early-type and stopped forming stars
at high redshift \cite{vd96,stanford98,vd01b}.

b) The blue galaxies observed in distant clusters, responsible for the
Butcher-Oemler effect, must largely have ``disappeared'' (= become
red) by z=0. \cite{kb01} (see also \cite{smail98}) 
have shown that the color-magnitude diagram
observed at intermediate z (rich of blue galaxies) can be reconciled
with the CM diagram of the Coma cluster (with very few luminous blue
galaxies) if star formation is halted in the blue galaxies at
intermediate redshifts.

c) Related to the previous point, some works at $z \sim 1$ argue for a
truncation of the CM sequence \cite{nakata01}, as only bright
galaxies are found to be in place on the sequence and there is a
deficit of fainter red galaxies. This is possibly related to the fact
that when star formation is halted in blue Butcher-Oemler galaxies they
evolve becoming redder {\sl and} fainter, going to populate the red
sequence at magnitudes fainter than the top brightest 1 or 2
magnitudes. Furthermore, a UV-excess has been found in the
galaxies on the red sequence as compared to passive evolution models
\cite{tanaka00,nakata01}.

d) On the color-magnitude sequence, there are also passive spirals at
all redshifts (\cite{p99} at z=0.5, \cite{couch01} at z=0.3,
\cite{terlevich01} in Coma, \cite{baloghnew} at z=0.25) and some 
mergers at z=0.8 \cite{vd99}.

All these findings suggest that the color-magnitude sequence of
clusters {\sl today} (except the very brightest end of it) must be
composed of a variegated population of galaxies that had a
variety of star formation histories. Though probably the 
mass-metallicity relation remains a main driver of the
CM relation, intricate age and metallicity
effects must be at work, as also shown by the age-metallicity
anticorrelation in passive galaxies found by a number of studies based
on spectral indices (see \cite{p01a} and references therein).
\vspace{-0.8cm}
\section{Spectroscopy: absorption-line spectra}
When the first spectra of galaxies in distant clusters
were taken, the biggest surprise 
was the finding of galaxies with strong Balmer lines in 
absorption and no emission lines. First recognized and named ``E+A''
by \cite{dg83}, their spectra were first modeled in detail by \cite{cs87}.
Now known also as ``k+a'' galaxies, they
can be explained if star formation was active in the recent past
and was halted at some point during the last 1-1.5 Gyr. The strongest
cases (equivalent width EW($\rm H\delta)> 5$ \AA) require a strong starburst
before the truncation of the star formation.
A large number of works have found and analyzed k+a spectra in distant clusters
\cite{dg83,cs87,henry87,fabri91,belloni95,barger96,pb96,fisher98,morris98,c98,abraham96}. The Mass-to-Light ratios of (the few studied)
k+a galaxies appear to be much lower than that of early-type galaxies, 
and are consistent with them having undergone a recent starburst 
\cite{kelson97,kelson00}.

In 10 clusters at z=0.4-0.5, the MORPHS collaboration
has found the k+a fraction to be significantly larger in clusters
than in the field at similar redshifts \cite{d99,p99}. These cluster k+a's 
have mostly spiral morphologies and present a radial distribution within
the cluster that is intermediate between the passive and the 
emission-line galaxy populations. 
In contrast, \cite{balogh99} interpret their results
on the CNOC clusters concluding there is no significant
excess of k+a's in the clusters at z=0.3 compared to the field.
The difference between the MORPHS and the CNOC results cannot
be ascribed to the optical vs X-ray cluster selection: the brightest 
X-ray clusters in both samples have similar X-ray luminosities, the
MORPHS clusters spanning a factor of 17 in X-ray luminosities while the
CNOC clusters a factor of 4, but the most X-ray luminous MORPHS
clusters tend to be those with the highest k+a fractions.
A Principal Component Analysis of the CNOC spectra by \cite{elli01}
finds a Post-Starformation component which again is intermediate in
dynamical state and stellar age between the old, passive population and the 
``field-like'' starforming component. This component presents a (small)
excess in clusters compared to the field, but the PCA cannot be easily 
translated into galaxy number fractions and thus be compared with the MORPHS
results.

An excess of k+a galaxies in clusters as opposed to the field is 
a strong evidence for a quenching of star formation in galaxies
as a consequence of the cluster environment. The implications
of k+a spectra regarding the presence of starburts in clusters
at high-z will be discussed in the next section. The fact that
k+a's mostly have spiral morphologies indicates that spectrophotometric
and morphological evolution are largely decoupled, and suggests
that the timescale for morphological transformation must be
longer than the k+a timescale ($> 1.5$ Gyr).
\vspace{-0.2cm}
\subsection{k+a galaxies at low redshift}
The occurrence of k+a galaxies at low z is believed to be low, both in
the field (0.1\%, \cite{zabludoff96}) and in clusters \cite{d87}. 
In Coma and other nearby clusters, Balmer strong
galaxies have been reported at a ``reduced frequency and burst
strength''(= weaker Balmer lines) compared to distant clusters
(\cite{cald93,cald97,rose01} and references therein). 
It has been suggested that in Coma this Balmer strong
galaxies are preferentially associated with the NGC4839 group
infalling from South-West \cite{cald93}, and the central
concentration of the latest star formation activity seems to support
the post-starburst scenario \cite{rose01}.  Recently, in a magnitude-limited
new spectroscopic survey of Coma galaxies, we find no k+a galaxy {\sl
as luminous as those in distant clusters}, but a significant
population of {faint} k+a's (dwarf galaxies, $M_V>-18.5$), with no
preference for the NGC4839 group \cite{p02}.  This work suggests that, while
strong-lined k+a spectra in luminous galaxies appear to be an
important phenomenon in distant clusters, in the local Universe the
k+a incidence is mostly related with dwarf galaxies.
\vspace{-0.8cm}
\section{Spectroscopy: emission-line spectra}
Emission lines are the most widely used 
indicators of ongoing star formation in the optical, 
because their flux is roughly proportional to the
current star formation rate. Surprisingly, the evolution with 
redshift of the fraction of cluster galaxies with emission lines
has not been properly quantified so far, although there is 
a trend for higher emission line fractions at high z (e.g. 
\cite{postman98,postman01}).
I try to show this graphically in Fig.~1, where I have plotted
the SFR per unit $L^{\star}$ luminosity (as derived from the [OII]3727
line) at z=0.5 and at z=0 for field galaxies (solid line)
and cluster galaxies (dotted line). At z=0.5 this is based on the MORPHS
cluster and field samples, and at z=0 on the Dressler \& Shectman
spectroscopic database (Dressler et al. in prep.). The most
striking result of this figure is the fact that the evolution in the 
clusters appears to be {\sl accelerated} with respect to the field,
in a similar way as found by \cite{kb01} 
on the basis of photometric data.
\begin{figure}
\vspace{-4cm}
\centerline{\includegraphics[width=24pc]{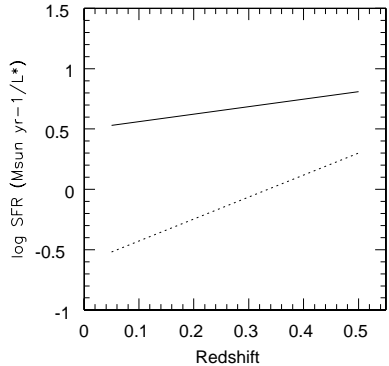}}
\vspace{-1.2cm}
\caption{Evolution of the SFR per unit $L^{\star}_B$ in the field (solid line) and cluster (dotted line).}
\end{figure}
Most works have instead focused on the clustercentric
radial behaviour of the emission line properties: the mean EW([OII])
is known to decrease with radius (e.g. \cite{bal97}). This mean EW
is calculated including all galaxies in the clusters (also early-type
galaxies) and in principle could be simply explained by the
morphology-density relation. However, even {\sl for a given morphological 
type}, the EW([OII]) distribution appears to be skewed towards
lower EWs in the cluster than in the field \cite{d99,bal98}.

A still open and debated question is whether, {\sl before} quenching
star formation, the cluster environment produces also a star formation
enhancement in the infalling galaxies.
Evidence for this enhancement arises from
the strong k+a cases, whose spectra can only be explained as post-starburst
galaxies. Considering that these strong-lined case must be the youngest 
(observed relatively soon after truncation),
and that soon they will evolve into k+a's with more moderate line strength,
the fraction of post-starburst galaxies among k+a spectra is necessarily
high. In principle, the starbursting progenitors of these post-starburst
galaxies could simply be field starburst galaxies that have infallen
into the clusters and had their star formation terminated. Whether
the starbursts in the field population are sufficient to account
for the k+a population observed in distant clusters is an issue
requiring further study (Dressler et al. in prep.).
Examples of cluster-induced starbursts have been observed in 
nearby clusters, where it is easier to study the star formation
signatures in great detail.

No need for a star formation enhancement in clusters is found
instead by looking at the [OII] or $\rm H\alpha$
equivalent width distributions
in clusters versus field and as a function of radius: no excess of 
emission line galaxies (as fractions of total number of
galaxies) is observed, as well as no tail at high
EWs (e.g. \cite{bal97,couch01}). 
However, eventual cluster-induced starbursts might have gone undetected
by this kind of analysis for at least three reasons: a) a possibly short
burst timescale; b) if the end-effect of the cluster is to quench
star formation, the fraction of emission-line 
galaxies detected does not say anything about
the amount of star formation ongoing in the (still) currently
starforming galaxies, and c) dust could moderate the line emission,
and it is expected to do so especially in those galaxies
with the highest star formation rates.
\vspace{-0.2cm}
\subsection{The role of dust}
There are several lines of evidence indicating the importance of
dust extinction in galaxies in distant clusters.
In the local Universe, galaxies with very strong emission lines
are usually {\sl faint} very late-type galaxies (Sd, Irr),
while massive and luminous starburst galaxies in most cases
show a combination of strong early Balmer lines in absorption and
weak-to-moderate [OII] emission (\cite{pw00} and references therein). 
These spectral
features are easily understood and reproduced by spectrophotometric
models in which younger stars suffer a higher dust extinction
than older stellar populations, as can be reasonably expected
\cite{p99,pbf01,bekki01,shioya00}.
Hence, searching for very strong emission lines {\tt is not}
an efficient method to identify starbursts with high SFRs.

Spectra resembling those of local massive starbursts with high dust
extinction have been found
in a significant fraction ($\sim 10$ \%) of both cluster and 
field galaxies at $z=0.5$ \cite{p99}.
Furthermore, the $\rm H\alpha$ line in emission has been detected
in some otherwise k+a or passive spectra, again suggesting
that the [OII] line in the blue might not have been detected due to dust
obscuration \cite{bal00,millernew}.

Additional evidence that SFRs can be strongly underestimated in the
optical comes from studies at dust-free wavelengths: radio continuum
emission has been detected in some of the strongest (=youngest) k+a
galaxies at z=0.4 \cite{smail99} and some optically passive galaxies
in low redshift clusters \cite{millernew}, suggesting they are
undergoing a current star formation activity that is invisible in the
optical. Mid-IR estimates of the SFR in star-forming galaxies in a
cluster at z=0.2 lead to values a factor 10 to 100 higher than those
based on the [OII] line \cite{duc02}, while standard dust
corrections usually adopted only account for a factor 2.5.

\vspace{-0.8cm}
\section{Morphologies}
Before HST, a few ground-based studies provided hints that the
galaxies responsible for the Butcher-Oemler effect are prevalently spirals
with disturbed morphologies, and/or mergers \cite{thompson86, 
lav88}. Already the first HST images proved that indeed
distant clusters contain a large number of spirals, many of which are
disturbed \cite{d94,c94,wirth94,d97,o97,c98}.
Moreover, in the MORPHS clusters, the high spiral fraction corresponds
to a low S0 fraction suggesting that a significant fraction of the spirals 
evolve into at least some of the S0s that dominate rich clusters today
(\cite{d97}, also \cite{kodama01}). This result has been questioned
by other studies \cite{andreon98,fabricant00}, but at least part of 
the discrepancy might be ascribed to the fact that these works have focused
on the number ratio of S0 to elliptical galaxies. This latter quantity,
more prone to errors than the simple fractions of each morphological type,
has been found to be highly dependent on the cluster characteristics,
in particular on the central concentration of ellipticals in
the clusters \cite{f00}. When considering separately the E, S0
and spiral fractions as a function of redshifts, studies at z=0.1-0.25 
confirm the trend of evolution of S0s \cite{f00}.
Also if the overall early-type fraction (E+S0) is examined, a
strong evolution with redshift is found \cite{vd00,lubinnew}
with high redshift clusters being composed of a much larger fraction
of late type galaxies.

Signatures for quite recent star formation in S0 galaxies consistent
with a transformation from spirals have been detected 
in several clusters, while
such activity is found to be absent in ellipticals
(\cite{thomas02,kunt98,p01b,smail01,vd98}, see also
Katgert in these proceedings). However, not all studies find a
difference among the stellar population ages of S0s and ellipticals
(e.g. \cite{ellis97, lewis01,ziegler01}). A possible explanation for
the discrepant results are the magnitude limits of the various
studies, since most of the S0s with recent star formation avoid the
top end of the luminosity function, as expected evolving the
luminosities of typical star-forming spirals at intermediate
redshifts \cite{p01b}.

Interestingly, cosmological high-resolution simulations are able to
account for the morphology-density relation of ellipticals, but
they fail to reproduce the properties of the S0s, as if some
additional mechanism for S0 formation, not currently included in
the models, might be required \cite{okamoto01,springel01}.

Finally, \cite{vd99} argue for an unexpectedly large fraction of mergers 
in a cluster at z=0.8. Since most of these pairs have red colors,
we might be witnessing the formation of ellipticals through the
merging of galaxies whose stars formed at high z,
as expected in hierarchical models of structure formation.

Due to space limitations, I am forced to omit in these proceedings
the discussion of the morphology-density relation \cite{d97,f00,thomas02,
dominguez01}, and of the ``star formation-density'' relation
\cite{lewis02,kodama01}.

\end{article}

\vspace{-0.8cm}
\begin{thebibliography}{200}
\bibitem{bo78} Butcher \& Oemler 1978, ApJ, 226, 559 + 1984, ApJ, 285, 426
\bibitem{smail98} Smail et al. 1998, MNRAS, 293, 124
\bibitem{andreon99} Andreon \& Ettori 1999, ApJ, 516, 647
\bibitem{kb01} Kodama \& Bower 2001, MNRAS, 321, 18 
\bibitem{elli01}Ellingson et al. 2001, ApJ, 547, 609
\bibitem{fairley02}Fairley et al. 2002, MNRAS, 330, 755
\bibitem{margoniner01}Margoniner et al. 2001, ApJ, 548, L143
\bibitem{metevier00}Metevier et al. 2000, AJ, 119, 1090 
\bibitem{wang97}Wang \& Ulmer 1997, MNRAS, 292, 920
\bibitem{pimbblet02}Pimbblet et al. 2002, MNRAS, 331, 333
\bibitem{abraham96}Abraham et al. 1996, ApJ, 471, 694
\bibitem{biviano97}Biviano et al. 1997, A\&A, 321, 84
\bibitem{moss00}Moss \& Whittle 2000, MNRAS, 317, 667
\bibitem{cald93}Caldwell et al. 1993, AnJ, 106, 473
\bibitem{cald97}Caldwell \& Rose 1997, AJ, 113, 492 
\bibitem{vd00}van Dokkum et al. 2000, ApJ, 541, 95
\bibitem{rakos95}Rakos \& Schombert 1995, ApJ, 439, 47
\bibitem{tanaka00}Tanaka et al. 2000, ApJ, 528, 123
\bibitem{ellis97}Ellis et al. 1997, ApJ, 483, 582
\bibitem{stanford98}Stanford et al. 1998, ApJ, 492, 461
\bibitem{kodama98}Kodama et al. 1998, A\&A, 334, 99
\bibitem{gladders98}Gladders et al. 1998, ApJ, 501, 571
\bibitem{bower92}Bower et al. 1992, MNRAS, 254, 601
\bibitem{stanford97}Stanford et al. 1997, AJ, 114, 2232 + 2002, AJ, 123, 619
\bibitem{rosati99}Rosati et al. 1999, AJ, 118, 76 
\bibitem{lubin00}Lubin et al. 2000, ApJ, 531, L5
\bibitem{kajisawa00}Kajisawa et al. 2000 PASJ, 52, 61
\bibitem{vd01a}van Dokkum et al. 2001a, ApJ, 552, L101
\bibitem{rosati01}Rosati 2001 private comm
\bibitem{postman98}Postman et al. 1998, AJ, 116, 560
\bibitem{p99}Poggianti et al. 1999, ApJ, 518, 576
\bibitem{vd96}van Dokkum \& Franx 1996, MNRAS, 281, 985
\bibitem{kelson97}Kelson et al. 1997, ApJ, 478, L13
\bibitem{kelson00}Kelson et al. 2000, ApJ, 531, 184 
\bibitem{kelson01}Kelson et al. 2001, ApJ, 552, L17
\bibitem{bender96}Bender et al. 1996, ApJ, 463, L51
\bibitem{ziegler97}Ziegler \& Bender 1997, MNRAS, 291, 527
\bibitem{ziegler01}Ziegler et al. 2001, MNRAS, 325, 1571
\bibitem{vd01b}van Dokkum et al. 2001b,, ApJ, 553, 90
\bibitem{nakata01}Nakata et al. 2001, PASJ 0110597
\bibitem{couch01}Couch et al. 2001, ApJ, 549, 820
\bibitem{terlevich01}Terlevich et al. 2001, MNRAS, 326, 1547
\bibitem{baloghnew}Balogh et al. astro-ph 0207360
\bibitem{vd99}van Dokkum et al. 1999, ApJ, 520, L95
\bibitem{p01a}Poggianti et al. 2001a, ApJ, 562, 689
\bibitem{dg83}Dressler \& Gunn 1983, ApJ, 270, 7 + 1992, ApJS, 78, 1
\bibitem{cs87}Couch \& Sharples 1987, MNRAS, 229, 423 
\bibitem{henry87}Henry \& Lavery 1987
\bibitem{fabri91}Fabricant et al. 1991, ApJ, 381, 33 + 1994, AJ, 107, 8
\bibitem{belloni95}Belloni et al. 1995, A\&A, 297, 61 + 1996, A\&AS, 118, 65
\bibitem{barger96}Barger et al. 1996, MNRAS, 279, 1
\bibitem{pb96}Poggianti \& Barbaro 1996, A\&A, 314, 379
\bibitem{fisher98}Fisher et al. 1998, ApJ, 498, 195
\bibitem{morris98}Morris et al. 1998, ApJ, 507, 84
\bibitem{c98}Couch et al. 1998, ApJ, 497, 188
\bibitem{d99}Dressler et al. 1999, ApJS, 122, 51
\bibitem{balogh99}Balogh et al. 1999, ApJ, 527, 54
\bibitem{zabludoff96}Zabludoff et al. 1996, ApJ, 466, 104
\bibitem{d87}Dressler 1987, in Nearly Normal Galaxies, Springer-Verlag, p.276
\bibitem{rose01}Rose et al. 2001, AJ, 121, 793
\bibitem{p02}Poggianti, Bridges, Carter, Mobasher et al. in prep. (see 0208181)
\bibitem{postman01} Postman et al. 2001, AnJ, 122, 1125
\bibitem{bal97}Balogh et al. 1997, ApJ, 488, L75
\bibitem{bal98}Balogh et al. 1998, ApJ, 504, L75
\bibitem{pw00}Poggianti \& Wu 2000, ApJ, 529, 157
\bibitem{pbf01}Poggianti, Bressan \& Franceschini 2001, ApJ, 550, 195
\bibitem{bekki01}Bekki et al. 2001, ApJ, 547, L17
\bibitem{shioya00}Shioya et al. 2000, ApJ, 539, L29 + 2001, ApJ, 558, 42
\bibitem{bal00}Balogh et al. 2000, ApJ, 540, 113 
\bibitem{millernew}Miller \& Owen 2002, astro-ph 0207662
\bibitem{smail99} Smail et al. 1999, ApJ, 525, 609
\bibitem{duc02}Duc et al. 2002, A\&A, 382, 60
\bibitem{thompson86}Thompson 1986, ApJ, 300, 639 + 1988, ApJ, 324, 112
\bibitem{lav88}Lavery \& Henry 1988, ApJ, 330, 596 + 1994, ApJ, 426, 524
\bibitem{d94}Dressler et al. 1994, ApJ, 430, 107
\bibitem{c94}Couch et al. 1994, ApJ, 430, 121
\bibitem{wirth94}Wirth et al. 1994, ApJ, 435, L105
\bibitem{d97}Dressler et al. 1997, ApJ, 490, 577
\bibitem{o97}Oemler et al. 1997, ApJ, 474, 561
\bibitem{andreon98}Andreon 1998, ApJ, 501, 533
\bibitem{fabricant00}Fabricant et al. 2000, ApJ, 539, 577
\bibitem{lubinnew}Lubin et al. astro-ph 0206442
\bibitem{f00}Fasano et al. 2000, ApJ, 542, 673 
\bibitem{thomas02}Thomas 2002, PhD Thesis, Leiden
\bibitem{kunt98}Kuntschner \& Davies 1998, MNRAS, 295, L29 
\bibitem{p01b}Poggianti et al. 2001b, ApJ, 563, 118
\bibitem{smail01}Smail et al. 2001, MNRAS, 323, 839 
\bibitem{vd98}van Dokkum et al. 1998, ApJ, 500, 714
\bibitem{lewis01}Lewis et al. 2001, ApJ, 528, 118
\bibitem{okamoto01}Okamoto \& Nagashima 2001, ApJ, 547, 109
\bibitem{springel01}Springel et al. 2001, MNRAS, 328, 726
\bibitem{dominguez01} Dominguez et al. 2001, AnJ, 121, 1266 + Balogh et al. 2002, ApJ, 566, 123
\bibitem{lewis02}Lewis et al. 2002, MNRAS, 334, 673 + Gomez, Nichol et al. 2002, in prep.
\bibitem{kodama01}Kodama et al. 2001, ApJ, 562, L9
\end{thebibliography}
\end{document}